\RequirePackage{fix-cm}
\documentclass{svjour3}                     % onecolumn (standard format)
\smartqed  % flush right qed marks, e.g. at end of proof

\usepackage{amsmath}
\usepackage{slashed}
\usepackage[utf8]{inputenc}
\usepackage{fontenc}
\usepackage{braket}
\usepackage{bbm}
\usepackage{cancel}
\usepackage{graphicx}
\usepackage[outdir=./]{epstopdf}
\usepackage{calrsfs}
\usepackage{emptypage}
\usepackage{fancyhdr}
\usepackage{nicefrac}
\usepackage{booktabs}
\usepackage{physics}
\usepackage{bbold}
\usepackage{color}
\usepackage{upgreek}
\newcommand{\1}{\mathbb{1}}
\renewcommand{\i}{\mathrm{i}}
\newcommand{\e}{\mathrm{e}}
\renewcommand{\d}{\mathrm{d}}

\begin{document}

\title{Entanglement in few-nucleon scattering events%\thanks{Grants or other notes
}


\author{Tanja Kirchner         \and
  Wael Elkamhawy \and
  Hans-Werner Hammer
}


\institute{Tanja Kirchner \at
           Technische Universität Darmstadt, Department of Physics, Institut f\"ur Kernphysik, 64289 Darmstadt, Germany\\
              \email{tkirchner@theorie.ikp.physik.tu-darmstadt.de}         %  \\
           \and
           Wael Elkamhawy  \at
           Technische Universität Darmstadt, Department of Physics, Institut f\"ur Kernphysik, 64289 Darmstadt, Germany\\
              \email{elkamhawy@theorie.ikp.physik.tu-darmstadt.de}           
              \and
              Hans-Werner Hammer \at
              Technische Universität Darmstadt, Department of Physics, Institut f\"ur Kernphysik, 64289 Darmstadt, Germany\\
              Helmholtz Forschungsakademie Hessen f\"ur FAIR (HFHF) and Extreme Matter Institute EMMI, GSI Helmholtzzentrum f\"{u}r Schwerionenforschung GmbH, 64291 Darmstadt, Germany\\
              \email{Hans-Werner.Hammer@physik.tu-darmstadt.de}
}

\date{Received: date / Accepted: date}

\maketitle

\begin{abstract}
  We investigate the spin entanglement in few-nucleon scattering
  processes involving nucleons and deuterons. For this purpose, we consider the entanglement power  
  introduced by Beane et al..
  We analyze different entanglement entropies as a  basis to define the 
  entanglement power of the strong interaction and
  calculate the corresponding entanglement powers for proton-neutron, neutron-deuteron, proton-deuteron, and deuteron-deuteron scattering. For the latter two processes, we also take into account the modification from the Coulomb interaction. In contrast to proton-neutron scattering, no universal low-energy features are
  evident in the spin entanglement in neutron-deuteron, proton-deuteron, and deuteron-deuteron scattering. 
\end{abstract}
\newpage
\section{Introduction}
\label{sec:intro}
Strongly interacting quantum systems can have universal properties that
are independent of their interaction at short distances~\cite{Braaten:2004rn}.
A well-known example is the low-energy scattering of bosons with
large $s$-wave scattering length $a$ and mass $m$.
If $a$ is positive and much larger than the range of the interaction $R$, there is a shallow 
two-body bound state with binding energy
$B_2 = 1/(m a^2)+{\cal O}(R/a)\,,$ 
and mean-square separation $a^2/2$. %\footnote{We use natural units with $\hbar=c=1$.}
If a third particle is added, a three-body parameter, $\kappa_*$, is required
to fully characterize the universal properties.
For fixed $a$, this implies universal correlations between 
different three-body observables parameterized by $\kappa_*$,
such as the Phillips line \cite{Phillips:1968zze}. Moreover, the
Efimov effect \cite{Efimov:1970zz} generates a universal spectrum
of three-body bound states characterized by $a$ and $\kappa_*$.

Universality is also manifest
in scattering observables. The scattering cross section of two particles
at energy $k^2/m$, e.g., takes the universal form
$d\sigma/d\Omega= 4a^2/(1+k^2 a^2)+{\cal O}(R/a)\,,$ and
becomes scale invariant in the unitary limit of infinite scattering length.
Similar universality relations exist for more bodies and more complicated
systems of hadrons and nuclei with spin and isospin degrees of freedom (see Refs.~\cite{Braaten:2004rn,Epelbaum:2008ga,Frederico:2012xh,Hammer:2017tjm,Hammer:2019poc,Kievsky:2021ghz} for more details). In few-nucleon systems, there is also an approximate Wigner $SU(4)$ symmetry that rotates the spin and isospin degrees 
of freedom into each other \cite{Mehen:1999qs,Bedaque:1999ve,Vanasse:2016umz}.

Methods from quantum information theory provide an alternative window
on the universal properties of strongly interacting quantum systems.
Beane et al. have shown that the suppression of spin entanglement
in the $S$-matrix for nucleon-nucleon scattering is correlated
with the Wigner $SU(4)$ spin-isospin symmetry~\cite{Beane:2018oxh}. Based on this
observation, they have conjectured that dynamical entanglement suppression
is a property of the strong interaction at low energies, giving
rise to Wigner $SU(4)$ as an emergent symmetry. This idea was further
elaborated in Ref.~\cite{Low:2021ufv} and extended to systems of
pions as well as pions and nucleons in Ref.~\cite{Beane:2021zvo}.
Bai and Ren presented a formalism able to account for the Coulomb
interaction using the screening method and investigated the entanglement entropy of $p-^3$He and
$n-^3$H scattering~\cite{Bai:2022hfv}.

In this work, we investigate the spin entanglement in the scattering
of spin-$\nicefrac{1}{2}$ and spin-$1$ particles with an application
to proton-neutron, nucleon-deuteron, and deuteron-deuteron scattering processes in mind. We consider different entanglement entropies as a  basis to define the 
entanglement power and investigate their
universal low-energy properties.
The entanglement of other degrees of freedom beyond spin is left for future work.
The case of spin-\nicefrac{1}{2}-spin-\nicefrac{1}{2} scattering based on the leading Taylor expansion of the von Neumann entropy
was already investigated by Beane et al.~\cite{Beane:2018oxh}.
The corresponding formalism for 
isospin-\nicefrac{1}{2}-isospin-1 and isospin-1-isospin-1 scattering was discussed in \cite{Beane:2021zvo}.
We use our results to calculate the entanglement powers for proton-neutron, nucleon-deuteron, and
deuteron-deuteron scattering.
For the proton-neutron and deuteron-deuteron cases, the modification from the Coulomb interaction is also taken into account. 

\section{Formalism}
\label{sec:formalism}
We follow Ref.~\cite{Beane:2018oxh} and consider two particles that are
initially uncorrelated. Thus, their spin state can be written as a product
of the separate one-particle spin states $\ket{\psi_\textrm{in}}=\ket{\psi}_1\otimes\ket{\psi}_2$.
We start by deriving a general initial state for spin-$\nicefrac{1}{2}$ and
spin-$1$ particles. Next, we analyze the correlation of the two spins induced by the scattering process.
For this purpose, we define the scattering operator $\hat{S}$,
i.e., the $S$-matrix, which transfers the initial state to the final scattered state
\begin{align}
\label{eq:defPsiout}
\ket{\psi_\textrm {out}}=\hat{S}\ket{\psi_\textrm{in}}\,.
\end{align}
The $S$-matrix is expressed  in terms of the spin operators and the phase shifts in the corresponding spin channels.

The final state $\ket{\psi_\textrm{out}}$ defines a density matrix
$\hat{\rho}=\ket{\psi_\textrm{out}}\bra{\psi_\textrm{out}}$ that contains
all quantum-mechanical information about the state. To calculate
correlations between the two final-state particles,
we need the reduced density matrix for particle 1
\begin{equation}
	\hat{\rho}_1=\Tr_2[\hat{\rho}]\,,
\end{equation}
where particle 2 has been traced out. 
Note that the labels $1$
and $2$ are arbitrary and our results do not depend on this choice. For 
definiteness, however, we will always 
assume that particle 2 has been traced out in the following.
Using the reduced density matrix, we can calculate entanglement entropies.
These entropies quantify the degree of ``entanglemen'' generated in the scattering process, which we will use as
a measure of correlation between the two particles. These entropies will be
shown in Sec.~\ref{sec:entropies}.

\subsection{Initial state}
\label{sec:initialState}
For our calculations, we consider a general product state of the pure one-particle spin states. This implies that the two particles are 
initially uncorrelated. For the pure one-particle spin states, 
we take an 
arbitrary vector on the corresponding state manifold \cite{bengtsson_zyczkowski_2006}.
This will give us every possible spin state that generates a suitable density 
matrix that is Hermitian, positive semidefinite, normalized to trace one
and is idempotent, i.e., \(\hat\rho^2=\hat\rho\). The spin states can be
parameterized by a set of angles. In the following, we will discuss the
general spin states for spin-$\nicefrac{1}{2}$ and spin-$1$, as these are the cases we
consider in our calculation.

\paragraph{Spin-$\nicefrac{1}{2}$}
For a spin-$\nicefrac{1}{2}$ particle, we have two orthogonal spin states:
$\ket{\nicefrac{1}{2},\nicefrac{1}{2}}$ and
$\ket{\nicefrac{1}{2},-\nicefrac{1}{2}}$. Therefore, two complex parameters (i.e., four real parameters) are required to specify a general pure state. 
By setting the overall phase and requiring the pure state to be properly normalized, 
two real parameters can be eliminated. The remaining two real parameters are angles 
that parameterize the corresponding complex manifold \textbf{CP}\(^1\) of all pure states
(projective Hilbert space of complex dimension 1). This space is also known as the \(2\)-sphere \textbf{S}\(^2\) or 
\textit{Bloch sphere}, which is the unit sphere in three dimensions. It can be parameterized as \cite{bengtsson_zyczkowski_2006}
\begin{equation}
    \label{eq:param1}
    \ket{\psi_{S=\nicefrac{1}{2}}}=\cos\vartheta_1\ket{\nicefrac{1}{2},\nicefrac{1}{2}}+\e^{\i\nu_1}\sin\vartheta_1\ket{\nicefrac{1}{2},-\nicefrac{1}{2}} \,,
\end{equation}
where $0<\vartheta_1<\pi/2$ and $0\leq\nu_1<2\pi$. The notion of distance of two points on \textbf{CP}\(^1\) 
is described by the \textit{Fubini-Study metric}
\begin{equation}
    \d s_{\text{FS}}^2(\textbf{CP}^1)=\d\vartheta_1^2+\frac{1}{4}
    \sin^2(2\vartheta_1)\d\nu_1^2\,.
\end{equation}
Since we need to average over all possible initial states for the entanglement power defined in Ref.~\cite{Beane:2018oxh}, we have to divide by the total volume of the corresponding manifold. 
In order to calculate the total volume of \textbf{CP}\(^1\), we require the differential \textit{Fubini-Study volume element}
\begin{equation}
    \d V_{\text{FS}}(\textbf{CP}^1) = \d \vartheta_1 \d\nu_1 \cos\vartheta_1 \sin\vartheta_1\,.
\end{equation}
Hence, the total volume reads
\begin{equation}
    V_{\text{FS}}(\textbf{CP}^1) = \int \d V_{\text{FS}}(\textbf{CP}^1) 
    = \pi\,.
\end{equation}
Since \textbf{CP}\(^1\) is isomorphic to \textbf{S}\(^2\), one could also use standard spherical coordinates in three dimensions and parameterize all pure states on the Bloch sphere \textbf{S}\(^2\) as
\begin{equation}
    \label{eq:param2}
    \ket{\psi_{S=\nicefrac{1}{2}}}=\cos(\frac{\vartheta}{2})\ket{\nicefrac{1}{2},\nicefrac{1}{2}}+\e^{\i\phi}\sin(\frac{\vartheta}{2})\ket{\nicefrac{1}{2},-\nicefrac{1}{2}} \,.
\end{equation}
Comparing the parameterizations in Eqs.~(\ref{eq:param1}) and (\ref{eq:param2}), we find \(\vartheta=2\vartheta_1\) and \(\phi=\nu_1\). The corresponding 
volume is \(4\pi\) instead of \(\pi\). 
However, averaging over all pure states will lead to the same results in either coordinates.

\paragraph{Spin-$1$}
For spin-$1$, there are three basis states given by the spin projections
$\ket{1,-1}$, $\ket{1,0}$ and $\ket{1,1}$.
An arbitrary state can be parameterized by three complex parameters (i.e., six real
parameters). Again, we eliminate two real parameters due to normalization and the overall phase. Therefore, we are left with four angles parameterizing the corresponding complex manifold \textbf{CP}\(^2\) of all pure 
states (projective Hilbert space of complex dimension 2). The resulting parameterization is given by \cite{bengtsson_zyczkowski_2006}\,
\begin{equation}
	\ket{\psi_{S=1}}=\cos\vartheta_1\sin\vartheta_2\ket{1,-1}+\e^{\i\nu_1}\sin\vartheta_1\sin\vartheta_2\ket{1,0}+\e^{\i\nu_2}\cos\vartheta_2\ket{1,1}\,,
\end{equation}
where $0<\vartheta_1,\vartheta_2<\pi/2$ and $0\leq\nu_1,\nu_2<2\pi$.
The differential volume element of the manifold is
\begin{equation}
	\d V_{\text{FS}}(\textbf{CP}^2)=\d\vartheta_1\,\d\vartheta_2\,\d\nu_1\,\d\nu_2 \cos\vartheta_1\cos\vartheta_2\sin\vartheta_1\sin^3\vartheta_2\,,
\end{equation}
and integration gives us the total volume
\begin{equation}
\begin{split}
  V_{\text{FS}}(\textbf{CP}^2)&= \int \d V_{\text{FS}}(\textbf{CP}^2)
  =\frac{\pi^2}{2}\,.
  \label{eq:norm1}
\end{split}
\end{equation}
Note that the corresponding ``generalized Bloch sphere'' for spin-1
is much more intricate than for spin-$\nicefrac{1}{2}$ and does not simply correspond to a 
unit sphere in 5 dimensions \cite{bengtsson_zyczkowski_2006,Goyal:2011xjg}.

\subsection{$S$-Matrix}
\label{sec:DerivationSMat}
Next, we derive a general  expression for the scattering operator.
We expand $\hat{S}$ in the identity matrix and powers of scalar
products of the one-particle spin operators. We use this ansatz up
to the square of the spin scalar product, as this is sufficient for spin-$\nicefrac{1}{2}$ and spin-1 degrees of freedom, 

\begin{equation}
	\hat{S}=%\sum_\alpha\,\,
	a	\,\1+
	b \,\vec{S}_1\cdot \vec{S}_2 
	+ c	\,(\vec{S}_1\cdot \vec{S}_2)^2\,,
	\label{eq:AnsatzSmat}
\end{equation}
where $\1$ is defined as the tensor product of the identity operators for the first and the second particle, i.e., $\1=\1_1\otimes\1_2$.
Similarly, the spin scalar product is defined by $\vec{S}_1\cdot\vec{S}_2=\sum_i S_1^i\otimes S_2^i$, where the sum runs over all Cartesian components of the
spin matrices for particle 1 and 2.
We determine the parameters in Eq.~\eqref{eq:AnsatzSmat} by demanding
\begin{equation}
	\mel{S=\alpha,M}{\hat{S}}{S=\alpha,M}=\e^{2\i\delta_\alpha}\,,
\end{equation}
for each spin channel $\alpha$ and all spin projections $M$.	
The scalar product of the spin matrices for particles 1 and 2 is expressed through squares of spin matrices in the usual way,
\begin{equation}
	\vec{S}_1\cdot\vec{S}_2=\frac{1}{2}\qty((\vec{S}_1+\vec{S}_2)^2-\vec{S}_1^2-\vec{S}_2^2)\,,
\end{equation}
which can be evaluated straightforwardly using eigenvalue relations.
A non-trivial check is given by the unitarity of $\hat{S}$. The condition $\hat{S}\hat{S}^\dagger=\hat{S}^\dagger\hat{S}=\1$ must hold true due to probability conservation.

\paragraph{Spin-$\nicefrac{1}{2}$-spin-$\nicefrac{1}{2}$}
As a test of our procedure, we rederive the expression for
$\hat{S}$ from \cite{Beane:2018oxh} for the $s$-wave scattering of spin-$\nicefrac{1}{2}$
nucleons. There are two spin channels,
$S=0$ and $S=1$, with phase shifts $\delta_0$ and $\delta_1$, respectively.
The ansatz from above leads to
the scattering operator \begin{equation}
\label{eq:S12-12}
  \hat{S}_{\frac{1}{2}\frac{1}{2}}=\frac{1}{4}\qty(\e^{2\i\delta_0}+3\e^{2\i\delta_1})\1-\qty(\e^{2\i\delta_0}-\e^{2\i\delta_1}){\vec{S}_1}\cdot{\vec{S}_2}\,,
\end{equation}
in agreement with the expression given in \cite{Beane:2018oxh}.

\paragraph{Spin-$1$-spin-$\nicefrac{1}{2}$}
In the $s$-wave scattering of a spin-$1$ and spin-$\nicefrac{1}{2}$ particle,
we have the two spin channels $S=\nicefrac{1}{2}$ and $S=\nicefrac{3}{2}$ with phase shifts $\delta_{1/2}$ and $\delta_{3/2}$, respectively.
Using the strategy described above, the scattering operator is found to be
\begin{align}
\label{eq:S12-2}
     \hat{S}_{1\frac{1}{2}}=\frac{1}{3}\qty(\e^{2\i\delta_{1/2}}+2\e^{2\i\delta_{3/2}})\1-\frac{2}{3}\qty(\e^{2\i\delta_{1/2}}-\e^{2\i\delta_{3/2}}){\vec{S}_1}\cdot{\vec{S}_2}\,,
\end{align}
which agrees with the results obtained in \cite{Beane:2021zvo} for the $\pi N$ system when spin operators are replaced by isospin operators.
     
\paragraph{Spin-$1$-spin-$1$}
For $s$-wave spin-$1$-spin-$1$ scattering, the total spins
$S=0$, $S=1$, and $S=2$ are possible. The general form of
the scattering operator for distinguishable particles is given by
\begin{align}
\label{eq:S1-1}
  \hat{S}_{11}=&
  -\frac{1}{3}\qty(\e^{2\i\delta_0}-3\e^{2\i\delta_1}-\e^{2\i\delta_2})\1
  -\frac{1}{2}\qty(\e^{2\i\delta_1}-\e^{2\i\delta_2}){\vec{S}_1}\cdot{\vec{S}_2}
  \nonumber\\
  &+\qty(\frac{1}{3}\e^{2\i\delta_0}-\frac{1}{2}\e^{2\i\delta_1}+\frac{1}{6}
  \e^{2\i\delta_2})({\vec{S}_1}\cdot{\vec{S}_2})^2\,,
\end{align}
with the corresponding phase shifts $\delta_{0}$, $\delta_{1}$, and $\delta_{2}$. Here, all three operators appearing in Eq.~\eqref{eq:AnsatzSmat} contribute. The equivalent result for isospin degrees of freedom was given in \cite{Beane:2021zvo} considering the case of $\pi\pi$ scattering. 

For $s$-wave scattering of two identical bosons with spin-1, only wave functions symmetric under particle exchange are allowed by Bose statistics.  Thus, only the total spins $S=0$ and $S=2$ are allowed. The $S=1$ state is forbidden by symmetry, i.e., neither interaction nor free propagation is allowed in this channel. Therefore, we construct an $S$-matrix that is unitary in the $S=0$ and $S=2$ channel but identically zero in the $S=1$ channel. This $S$-matrix can be obtained from 
Eq.~(\ref {eq:S1-1}) by replacing $\e^{2\i\delta_1}\to 0$,
which leads to the scattering operator 
\begin{align}
\label{eq:S1-1_2}
  \hat{S}_{dd}=&
  -\frac{1}{3}\qty(\e^{2\i\delta_0}-\e^{2\i\delta_2})\,\1
  +\frac{1}{2}\e^{2\i\delta_2}\,{\vec{S}_1}\cdot{\vec{S}_2}
  \nonumber\\
  &+\frac{1}{6}\qty(2\e^{2\i\delta_0}+
  \e^{2\i\delta_2})\,({\vec{S}_1}\cdot{\vec{S}_2})^2\,.
\end{align}
Formally, this corresponds to applying a projection operator $\hat{P}$ to $S_{11}$ which projects the $S$-matrix on the subspace with $S=0,2$,
\begin{align}
\label{eq:defP}
   \hat{S}_{dd}= \hat{P}^\dagger \hat{S}_{11} \hat{P}\,. 
\end{align}
Alternatively, the operator $\hat{P}$ can also act on the initial product state $\ket{\psi_\textrm{in}}$ in Eq.~\eqref{eq:defPsiout}, leading to the same result for the
entanglement entropy. We will come back to this viewpoint below when we discuss the entanglement power for $dd$ scattering.

\subsection{Entropies}
\label{sec:entropies}

To quantify the correlation between the two spins after the scattering process, we use entropies as a measure of the entanglement. We examine different entropy definitions as a basis for the entanglement power 
and investigate whether they are equally suitable to describe the generated entanglement.

First, we consider the standard von Neumann entropy
\begin{equation}
E_N(\hat{\rho}_1)=-\Tr[\hat{\rho}_1\ln(\hat{\rho}_1)]=-\sum_i \lambda_i\ln(\lambda_i)\,,
\end{equation}
where the $\lambda_i$ with $i=1,2,\ldots$ denote the eigenvalues
of the reduced density matrix $\hat{\rho}_1$.
Since the calculation of the von Neumann entropy requires the diagonalization
of the density matrix, it is often convenient to study the 
Taylor expansions of the von Neumann entropy of order $n$,
\begin{equation}
\label{eq:vN}
E_n(\hat{\rho}_1)=-\Tr[\hat{\rho}_1\,\sum_{j=1}^n\frac{(-1)^{j+1}\,(\hat{\rho}_1-\1)^j}{j}]\,.
\end{equation}
Finally, we also investigate the Rényi entropies
\begin{align}
\label{eq:Renyi}
    E_R(\hat\rho_1,\alpha)=&\frac{1}{1-\alpha}\ln\Big(\Tr[\hat\rho_1^\alpha]\Big), \qquad \alpha>0\,,\\
    =&\frac{1}{1-\alpha}\ln(\sum_i \lambda_i^\alpha)\,,
\end{align}
for $\alpha=0.5$ and $\alpha=2$. Note that in the limit $\alpha\to 1$,
the Rényi entropy tends to the von Neumann entropy.

\subsection{Averaging over initial states: entanglement power}
Following \cite{Beane:2018oxh}, the ``entanglement power'' of the $S$-Matrix 
is defined by calculating the entropy $E$ of particle 1 for a reduced density matrix
$\hat{\rho}_1=\Tr_2[\hat{\rho}]\,,$ and
averaging the entropy $E$ over all possible initial states of the
scattering process, meaning
\begin{equation}
\label{eq:entang_power}
\epsilon= \frac{1}{V_{\text{FS(1)}} V_{\text{FS(2)}}} \int \d V_{\text{FS(1)}} \d V_{\text{FS(2)}} \,E\,,
\end{equation}
where $\d V_{\text{FS(1)}}$ and $\d V_{\text{FS(2)}}$ denote the Fubini-Study volume elements for particle 1 and 2, respectively.
This average cancels out the dependence on the initial state. For the 
leading Taylor expansion of the von Neumann entropy and spin-$\nicefrac{1}{2}$
particles, this reduces to the definition of Ref.~\cite{Beane:2018oxh},
\begin{equation}
\epsilon= 1-\frac{1}{16\pi^2 }\int \d \Omega_1 \d \Omega_2 \Tr_1 [\hat\rho_1^2]\,.
\end{equation}
Next, we revisit the case of nucleon-nucleon scattering and
apply our results to experimental data for nucleon-deuteron and
deuteron-deuteron scattering.

\section{Application to nuclear scattering processes}
\subsection{Neutron-proton scattering}
 
We start with the case of neutron-proton scattering, which was already discussed in \cite{Beane:2018oxh} for the entanglement power based on the leading Taylor 
expansion of the von Neumann entropy.
Using the form of the $S$-matrix, Eq.~\eqref{eq:S12-12}, and
the expressions in Sec.~\ref{sec:entropies},
the entanglement powers can be calculated from the scattering 
phase shifts.  Since we are interested in scattering close to threshold, we focus on the $s$-wave contribution. 

The Taylor expansions $E_n$ of the Neumann entropy as function of the reduced density matrix $\hat\rho$, Eq.~\eqref{eq:vN}, up to order $n=7$ are listed in Table~\ref{tab:general_taylor}. Their calculation does not require the diagonalization of the density matrix.
%%%%%%%%%%%%%%%%%%%%%%%%%%%%%%%%%%%%%%%%%%%%%%%%%%%%%%%%%%%%%%%%%%%%%%%%%%%
\begin{table}[h]
    \caption{Taylor expansions $E_n$ of the von Neumann entropy $E_N$ as function of the reduced density matrix $\hat\rho$ up to order $n=7$.}
    \centering
    \begin{tabular}{cl}
         \toprule
         $n$ & $E_n$\\
         \midrule
         1& $1-\Tr[\hat{\rho}^2]$\\[.5ex]
         2& $\frac{3}{2}-2\Tr[\hat{\rho}^2]+\frac{1}{2}\Tr[\hat{\rho}^3]$\\[.5ex]
         3&  $\frac{11}{6}-3\Tr[\hat{\rho}^2]+\frac{3}{2}\Tr[\hat{\rho}^3]-\frac{1}{3}\Tr[\hat{\rho}^4]$\\[.5ex]
         4&  $\frac{25}{12}-4\Tr[\hat{\rho}^2]+3\Tr[\hat{\rho}^3]-\frac{4}{3}\Tr[\hat{\rho}^4]+\frac{1}{4}\Tr[\hat{\rho}^5]$\\[.5ex]
         5 &  $\frac{137}{60}-5\Tr[\hat{\rho}^2]+5\Tr[\hat{\rho}^3]-\frac{10}{3}\Tr[\hat{\rho}^4]+\frac{5}{4}\Tr[\hat{\rho}^5]-\frac{1}{5}\Tr[\hat{\rho}^6]$\\[.5ex]
         6 &  $\frac{49}{20}-6\Tr[\hat{\rho}^2]+\frac{15}{2}\Tr[\hat{\rho}^3]-\frac{20}{3}\Tr[\hat{\rho}^4]+\frac{15}{4}\Tr[\hat{\rho}^5]-\frac{6}{5}\Tr[\hat{\rho}^6]+\frac{1}{6}\Tr[\hat{\rho}^7]$\\[.5ex]
         7 &  $\frac{363}{140}-7\Tr[\hat{\rho}^2]+\frac{21}{2}\Tr[\hat{\rho}^3]-\frac{35}{3}\Tr[\hat{\rho}^4]+\frac{35}{4}\Tr[\hat{\rho}^5]-\frac{21}{5}\Tr[\hat{\rho}^6]
         +\frac{7}{6}\Tr[\hat{\rho}^7]-\frac{1}{7}\Tr[\hat{\rho}^8]$\\[.5ex]
         \bottomrule
    \end{tabular}
    \label{tab:general_taylor}
\end{table}
%%%%%%%%%%%%%%%%%%%%%%%%%%%%%%%%%%%%%%%%%%%%%%%%%%%%%%%%%%%%%%%%%%%%%%%%%%%
Analytic expressions for the $s$-wave contribution to the corresponding entanglement powers $\epsilon_n$ in terms of the spin-singlet and spin-triplet phase shifts $\delta_0$ and $\delta_1$ are therefore straightforward to calculate.
Inserting the general form of the $S$-matrix, Eq.~\eqref{eq:S12-12},
we obtain the expressions for $\epsilon_n$ given in Table~\ref{tab:pn_taylor}.
%%%%%%%%%%%%%%%%%%%%%%%%%%%%%%%%%%%%%%%%%%%%%%%%%%%%%%%%%%%%%%%%%%%%%%%%%%%
\begin{table}[h]
    \caption{$s$-wave scattering contributions to the entanglement powers $\epsilon_n$ based on Taylor expansions $E_n$ of the von Neumann entropy $E_N$ for $n=1\ldots 7$ expressed through the spin-singlet and spin-triplet phase shifts $\delta_0$ and $\delta_1$.
    }
    \centering
    \begin{tabular}{c|l}
    \toprule
         $n$ &
         $\epsilon_n$\\[.5ex]
         \midrule
         1 & $\frac{1}{6}\sin[2(\delta_0-\delta_1)]^2$ \\[.5ex]
         2 & $\frac{5}{24}\sin[2(\delta_0-\delta_1)]^2$\\[.5ex]
         3 & $\frac{1}{720}\qty(167+3\cos[4(\delta_0-\delta_1)])\sin[2(\delta_0-\delta_1)]^2$\\[.5ex]
         4 & $\frac{1}{5760}\qty(1429+51\cos[4(\delta_0-\delta_1)])\sin[2(\delta_0-\delta_1)]^2$\\[.5ex]
         5 & $\frac{1}{403200}\qty(104869+5406\cos[4(\delta_0-\delta_1)]+45\cos[8(\delta_0-\delta_1)])\sin[2(\delta_0-\delta_1)]^2$\\[.5ex]
         6 & $\frac{1}{537600}\qty(144867 + 9508 \cos[4 (\delta_0 - \delta_1)] + 185\cos[8 (\delta_0 - \delta_1)]) \sin[2 (\delta_0 - \delta_1)]^2$\\[.5ex]
         7 & $\frac{1}{45158400}(12511358 + 978417 \cos[4 (\delta_0 - \delta_1)] + 
   30690 \cos[8 (\delta_0 - \delta_1)]$\\[.5ex]
    & $+175 \cos[12 (\delta_0 - \delta_1)]) \sin[2 (\delta_0 - \delta_1)]^2$\\
    \bottomrule
    \end{tabular}
    \label{tab:pn_taylor}
\end{table} 
%%%%%%%%%%%%%%%%%%%%%%%%%%%%%%%%%%%%%%%%%%%%%%%%%%%%%%%%%%%%%%%%%%%%%%%%%%%
The entanglement powers based on the von Neumann entropy, $\epsilon_N$, and Rényi entropies, $\epsilon_R$, are calculated numerically.

\begin{figure}[h]
    \centering
    \includegraphics[width=.49\textwidth]{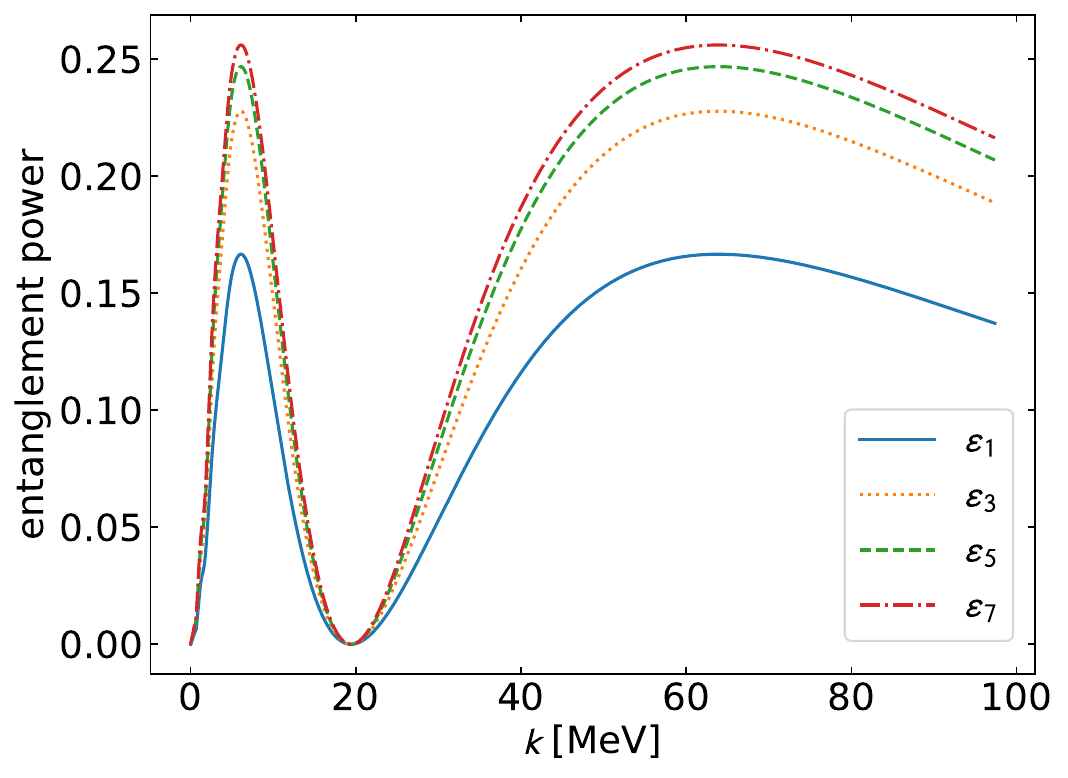}
    \includegraphics[width=.49\textwidth]{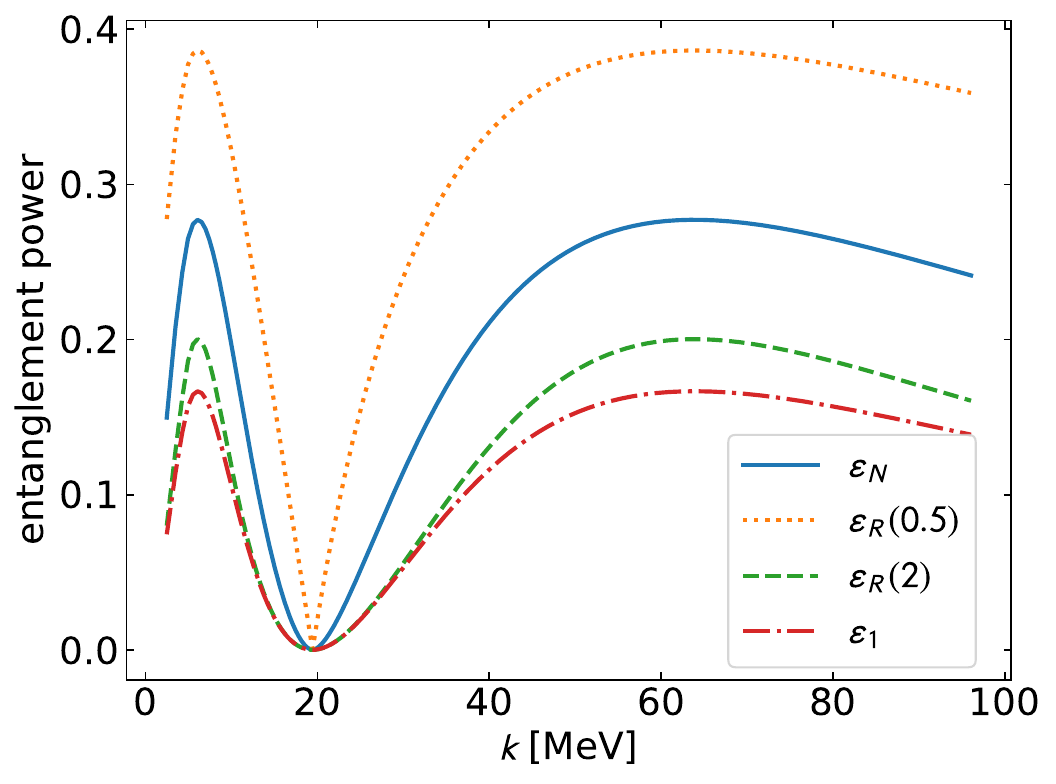}
    \caption{Comparison of the different entanglement powers as a function of the relative momentum $k$ evaluated using phase shift data from the Nijmegen partial wave analysis PWA93 \cite{Stoks:1993tb,pwa93} for $pn$-scattering. \textit{Left panel}: using different Taylor expansions of the von Neumann entropy: $\epsilon_i$, $i=1,3,5,7$.  \textit{Right panel}: using the von Neumann entropy $\epsilon_N$ and Rényi entropies $\epsilon_R$ for $\alpha=0.5$ and $\alpha=2$ compared to $\epsilon_1$.}
    \label{fig:pn_protneut_taylorexp}
\end{figure}

The entanglement powers can be evaluated using the Nijmegen partial wave analysis PWA93 \cite{Stoks:1993tb,pwa93}. Other potential models have been evaluated in \cite{Beane:2018oxh}
and lead to similar results.
The evaluation of the entanglement powers based on the Taylor-expanded von Neumann entropy, $\epsilon_n$ for $n=1,3,5,7$, with phase shift data from the Nijmegen partial wave analysis PWA93 \cite{Stoks:1993tb,pwa93} are shown in the left plot of Fig.~\ref{fig:pn_protneut_taylorexp}. 
In the right panel, we show a comparison of $\epsilon_1$ with the entanglement power based on the full von Neumann entropy, $\epsilon_N$, and the Rényi entropies $\epsilon_R$
for $\alpha=0.5$ and $\alpha=2$.
%%%%%%%%%%%%%%%%%%%%%%%%%%%%%%%%%%%%%%%%%%%%%%%%%%%%%%
%%%%%%%%%%%%%%%%%%%%%%%%%%%%%%%%%%%%%%%%%%%%%%%%%%%%%%
Evidently, the qualitative behavior of all entanglement powers is the same, they are just scaled differently. In particular, the position of minima and maxima stays the same. Thus, all definitions carry the same physical information. We observe that the difference between neighboring Taylor expansions $\epsilon_n$ decreases with $n$. This suggests that the expansion in Eq.~\eqref{eq:vN} converges for $pn$ scattering  and all entanglement powers are equally suited to describe the entanglement created in the scattering process. Thus, our study confirms the results of Ref.~\cite{Beane:2018oxh} for the $pn$ case, including the evidence for an emergent Wigner $SU(4)$ symmetry. 

\subsection{Phenomenology}
The qualitative behavior in the $pn$ case can also be understood based on
the effective range expansion of the phase shifts, % for $s$-waves ($l=0$),
\begin{equation}
\label{eq:ere-exp}
	k\cot\delta_j(k)=-\frac{1}{a_j}+\frac{r_j}{2}k^2+\dots\,,
	\quad j=0,1\,,
\end{equation}
where $k$ is the relative momentum and the total energy in the center of mass frame is $k^2/m$. Moreover, $a_j$ and $r_j$ are the scattering length and effective range in the spin channel $j$, respectively.

Since, in our case, the scattering lengths are large, we can neglect effective range effects 
at low energies, $kr_j\ll 1$, and the phase shifts can be approximated 
by $\delta_j(k)=\arccot(-1/(a_j k))$. Inserting this expression for both channels in the entanglement power
$\epsilon_1$ from Table~\ref{tab:pn_taylor}
gives us
\begin{equation}
	\epsilon_1=\frac{2}{3}\frac{(a_1-a_0)^2k^2(1+a_0 a_1 k^2)^2}{(1+a_0^2k^2)^2(1+a_1^2k^2)^2},
\end{equation}
which allows us to determine the minima and maxima analytically. The minima of $\epsilon_1$ are at $k_\mathrm{min,1}=0$ MeV and $k_\mathrm{min,2}=(-a_1 a_0)^{-1/2}$, while
the maxima are given by
\begin{equation}
\	k_\mathrm{max,\pm}=\frac{\pm(a_0-a_1)-\sqrt{a_0^2-6a_0 a_1+a_1^2}}{2a_0 a_1}.
\end{equation} 
Using the explicit values $a_0=5\,\mathrm{fm}$ and $a_1=-20\,\mathrm{fm}$ gives the minima at $k_\mathrm{min,1}=0~\mathrm{MeV}$ and $k_\mathrm{min,2}\approx 20\,\mathrm{MeV}$ and the maxima at $k_\mathrm{max,+}\approx7\,\mathrm{MeV}$ and $k_\mathrm{max,-}\approx57\,\mathrm{MeV}$.
This agrees well with the full numerical results shown in Fig.~\ref{fig:pn_protneut_taylorexp}.

\section{$nd$ and $pd$ scattering}
%	\subsection{scattering matrix}
Restricting ourselves to $s$-wave scattering, the nucleon and the deuteron can scatter in the $^2S_{1/2}$ and $^4S_{3/2}$ channels. The general form of the scattering matrix for this case is given in Eq.~\eqref{eq:S12-2}.

\subsection{Coulomb interaction}
\label{sec:ndCoulomb}
In the $pd$ case, we also need to take into account the Coulomb interaction between the proton and the deuteron.
The $s$-wave phase shifts in presence of a strong interaction and a Coulomb interaction split up into three pieces,
\begin{equation}
	\delta_\mathrm{tot}(k)=-\eta_k \log(2 k r)
	+\sigma(k)+\delta_N(k)\,,
	\label{eq:Cb-phase}
\end{equation}
a logarithmic part, a pure Coulomb part $\sigma$, and a Coulomb-modified nuclear part $\delta_N$ (also known as Coulomb-subtracted phase shift), see, e.g., Refs.~\cite{Taylor:1972pty,HigaRupak_CoulombTreatment}.
Here, $\delta_N$ is the additional strong phase shift relative to the Coulomb wave function while the pure $s$-wave Coulomb phase shift $\sigma$ is given by $\sigma(k)=\arg \Gamma(1+\i\eta_k)$. Moreover,
$\eta_k=\alpha_e Z_1 Z_2\mu/k$ is the Sommerfeld parameter, $k$ is the relative momentum of the scattered particles, $\alpha_e=e^2/(4\pi)$ is the electromagnetic fine structure constant in Heaviside-Lorentz units, $\mu$ is their reduced mass, and the $Z_i$ are their charge numbers. Defining the Coulomb momentum scale $k_c=\alpha_e Z_1 Z_2 \mu$, the Sommerfeld parameter can also be written as $\eta_k=k_c/k$.
The amplitude describing the effects of the strong force
relative to the Coulomb interaction takes the form
\begin{equation}
    f_{SC}(k)=\frac{e^{2i\sigma(k)}}{k\cot\delta_N(k)-ik}\,,
\end{equation}
where the pure Coulomb phase shift enters as a prefactor. 
We can set up our scattering matrix the same way as before, but this time with the Coulomb-subtracted phase shift $\delta_N$. We emphasize that $\delta_N$ is defined relative to the outgoing Coulomb waves instead of plane waves as in the case without Coulomb.

We will see below that the entanglement entropies for $pd$ scattering only depend on the difference of the $s$-wave phase shifts $\delta_{1/2}$ and $\delta_{3/2}$.
Thus, our procedure can be justified 
by considering screened Coulomb potentials as in Ref.~\cite{Bai:2022hfv}. 
The contribution from the first two terms in Eq.~(\ref{eq:Cb-phase}) will simply cancel out in the entanglement entropy, such that the screening can safely be removed.
As a consequence, the effect of Coulomb interaction on the entanglement enters only via the Coulomb-modified nuclear part $\delta_N$. It can only be observed in the difference between $nd$ and $pd$ scattering data.
This result is not unexpected since the Coulomb interaction is spin-independent and thus cannot create any spin entanglement.

\subsection{Results}
\begin{figure}[h!]
    \centering
    \includegraphics[width=.45\textwidth]{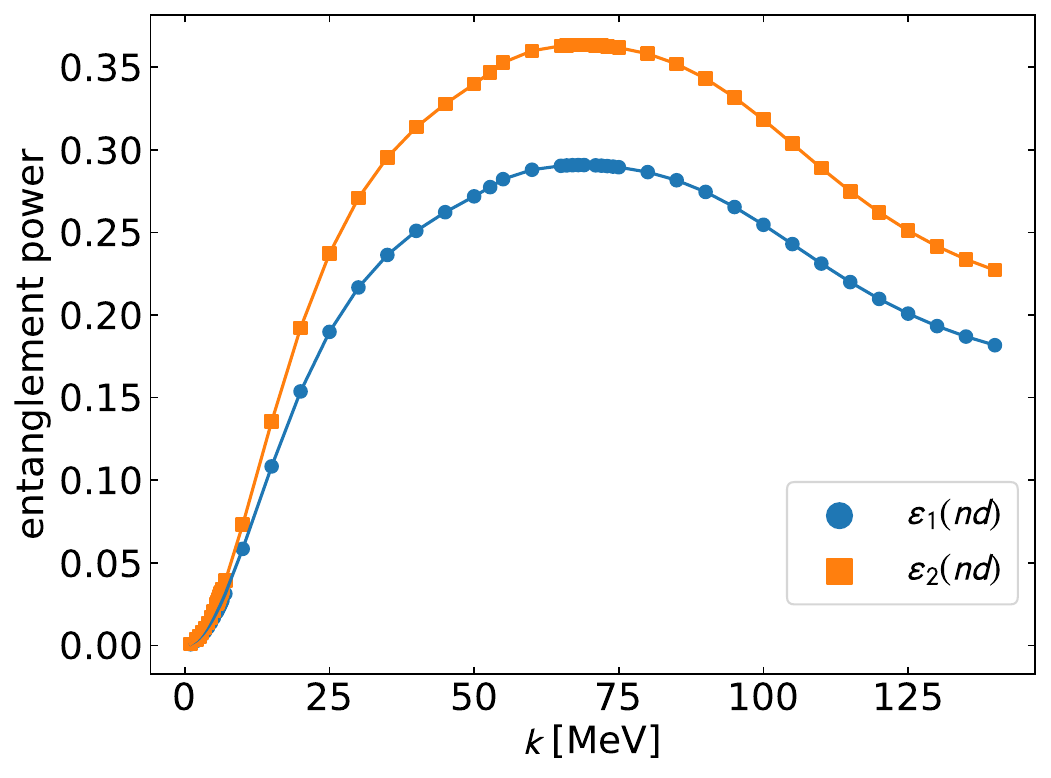}
    \includegraphics[width=.45\textwidth]{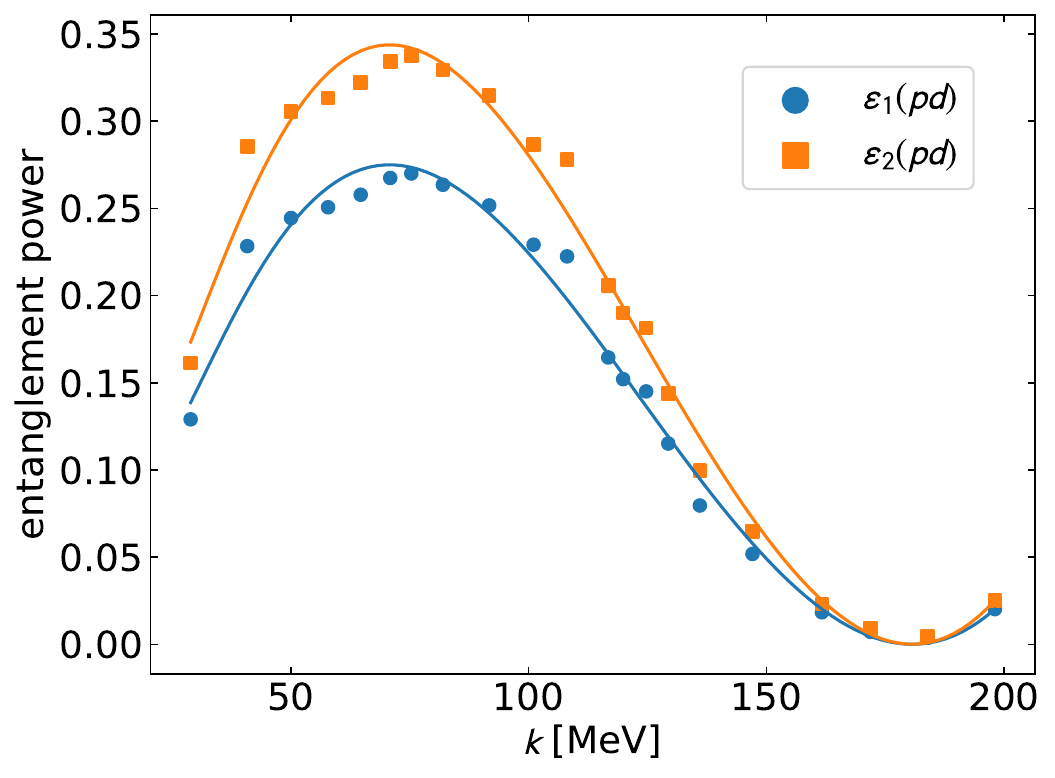}
    \caption{Comparison of the entanglement powers $\epsilon_1$ and $\epsilon_2$ for $nd$ and $pd$ scattering based on the Taylor expansions of the von Neumann entropy. \textit{Left panel}: $nd$ scattering with phase shifts from the pionless effective field theory calculation of Vanasse~\cite{nD_phaseshifts_Vanasse}. \textit{Right panel}: $pd$ scattering based on phase shifts obtained from experimental data analyzed in Ref.~\cite{Arvieux:1974fma}.
    In both plots the data are represented by points while the curves are inserted to guide the eye.
    }
    \label{fig:nd_taylor}
\end{figure}

We have calculated analytical expressions for the entanglement power based on the first two Taylor expansions of the von Neumann entropy. They are given in Table~\ref{tab:nd_taylor}. The result for the first Taylor expansion, $\epsilon_1$, agrees with the entanglement entropy obtained in \cite{Beane:2021zvo} for $\pi N$ scattering. The entanglement powers based on the von Neumann entropy, $\epsilon_N$, and Rényi entropies, $\epsilon_R$, for $\alpha=0.5$ and $\alpha=2$ will be calculated numerically as before.
%%%%%%%%%%%%%%%%%%%%%%%%%%%%%%%%%%%%%%%%%%%%%%%%%%%%%%%%%%%%%%%%%%%%%%%%%%%%%%%%%%%%%%%%%%%%%%%%%%
\begin{table}[b]
    \caption{$s$-wave scattering contributions to the entanglement powers $\epsilon_n$ based on Taylor expansions $E_n$ of the von Neumann entropy $E_N$ for $n=1,2$ expressed through the spin-doublet and spin-quartet phase shifts $\delta_{1/2}$ and $\delta_{3/2}$. 
    }
    \centering
    \begin{tabular}{c|l}
    \toprule
         $n$ & $\epsilon_n$  \\
         \midrule
         1 & $\frac{8}{243}\qty(17+10\cos[2(\delta_{1/2}-\delta_{3/2})])\sin[\delta_{1/2}-\delta_{3/2}]^2$ \\[.5ex]
         2 & $\frac{10}{243}\qty(17+10\cos[2(\delta_{1/2}-\delta_{3/2})])\sin[\delta_{1/2}-\delta_{3/2}]^2$\\
         \bottomrule
    \end{tabular}
    \label{tab:nd_taylor}
\end{table}
%%%%%%%%%%%%%%%%%%%%%%%%%%%%%%%%%%%%%%%%%%%%%%%%%%%%%%%%%%%%%%%%%%%%%%%%%%%%%%%%%%%%%%%%%%%%%%%%%%%%
One can already see that the two expansions only differ by a constant rescaling. Evaluating these expressions for $nd$ and $pd$ scattering data gives us the left and right plot in Fig.~\ref{fig:nd_taylor}, respectively.
As discussed above, only the Coulomb-modified nuclear phase shift $\delta_N$
contributes while the pure Coulomb contribution, which is the same in both spin channels, cancels out.
%%%%%%%%%%%%%%%%%%%%%%%%%%%%%%%%%%%%%%%%%%%%%%%%%%%%%%%%%%%%%%%%%%%%%%%%%%%%%%%%%%%%%%%%%%%%%%%%%%%%
%%%%%%%%%%%%%%%%%%%%%%%%%%%%%%%%%%%%%%%%%%%%%%%%%%%%%%%%%%%%%%%%%%%%%%%%%%%%%%%%%%%%%%%%%%%%%%%%%%%
In the left panel, we compare $\epsilon_1$ and $\epsilon_2$ for $nd$ scattering using phase shifts from the pionless effective field theory calculation of Vanasse~\cite{nD_phaseshifts_Vanasse}.
In the right panel, we show the corresponding results for $pd$ scattering based on phase shifts obtained from experimental data analyzed in Ref.~\cite{Arvieux:1974fma}. Both plots show similar qualitative
behavior with a minimum at relatively high momenta of order 150 MeV. This minimum is not expected to be governed by universal low-energy physics.

%%%%%%%%%%%%%%%%%%%%%%%%%%%%%%%%%%%%%%%%%%%%%%%%%%%%%%%%%%%%%%%%%%%%%%%%%%%%%%%%%%%%%%%%%%%%%%%%%%%%
\begin{figure}[h!]
    \centering
    \includegraphics[width=.65\textwidth]{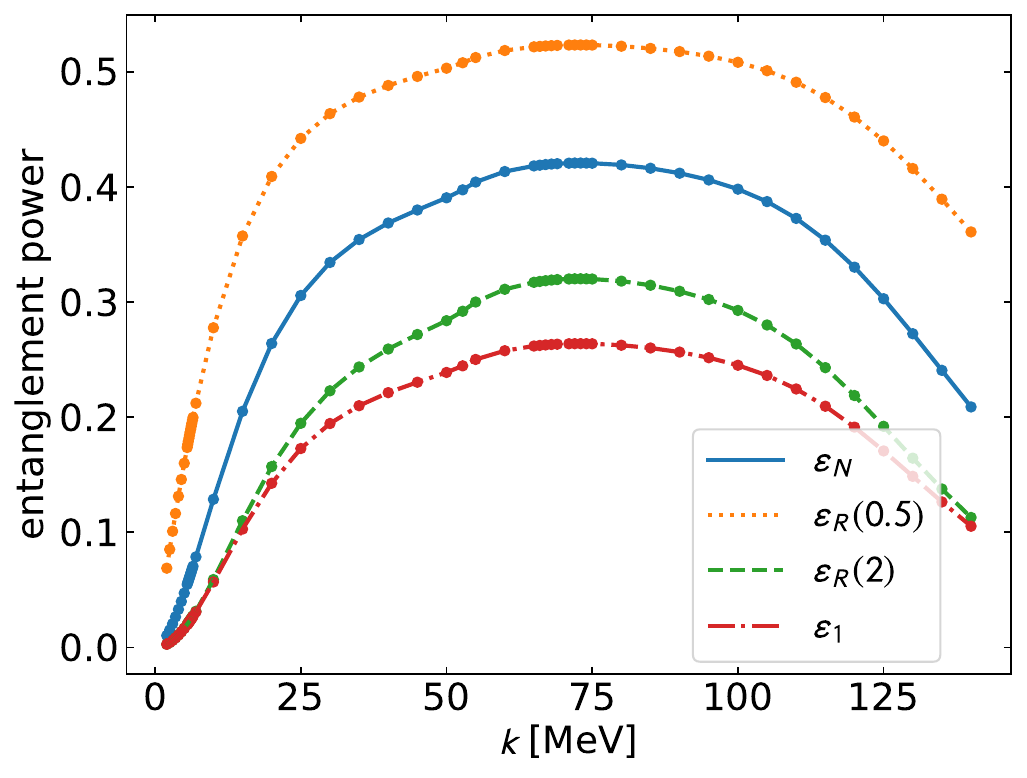}
    \caption{Comparison of the entanglement powers for $nd$ scattering based on the von Neumann entropy, $\epsilon_N$, and Rényi entropies, $\epsilon_R$, for $\alpha=0.5$ and $\alpha=2$ with $\epsilon_1$. Phase shifts are taken from the pionless effective field theory calculation of Vanasse~\cite{nD_phaseshifts_Vanasse}. The data are represented by points. The lines are added to guide the eye.}
    \label{fig:nd_entropy_comp}
\end{figure}
%%%%%%%%%%%%%%%%%%%%%%%%%%%%%%%%%%%%%%%%%%%%%%%%%%%%%%%%%%%%%%%%%%%%%%%%%%%%%%%%%%%%%%%%%%%%%%%%%%%
Figure~\ref{fig:nd_entropy_comp} shows the comparison of entanglement powers for $nd$ scattering based on different entropies as discussed in the caption. As in the nucleon-nucleon case, the qualitative features are very similar such that
the much easier to calculate entanglement powers $\epsilon_1$ and $\epsilon_2$ are sufficient for our purposes. Note also that no characteristic signature of the triton virtual state below the scattering 
threshold \cite{Reiner:1969mmv,Phillips:1969hm,Rupak:2018gnc}
can be seen since the entanglement power vanishes at $k=0$.

\subsection{Phenomenology}
Inserting the effective range expansion for $s$-waves given in Eq.~\eqref{eq:ere-exp}
for the doublet and quartet channels into the 
analytical expression for $\epsilon_1$ from Table~\ref{tab:nd_taylor}
allows for a discussion 
of the qualitative features based on effective range parameters.
Keeping only the scattering length contribution to the phase shifts, we obtain
\begin{eqnarray}
	\epsilon_1 &= &\frac{8}{243}\qty(17+10\frac{1-(a_{\nicefrac{1}{2}}^2-4a_{\nicefrac{1}{2}} a_{\nicefrac{3}{2}}+a_{\nicefrac{3}{2}}^2)k^2+a_{\nicefrac{1}{2}}^2 a_{\nicefrac{3}{2}}^2k^4}{(1+a_{\nicefrac{1}{2}}^2k^2)(1+a_{\nicefrac{3}{2}}^2k^2)})\nonumber\\
	&&\times
	\frac{(a_{\nicefrac{1}{2}}- a_{\nicefrac{3}{2}})^2k^2}{(1+a_{\nicefrac{1}{2}}^2k^2)(1+a_{\nicefrac{3}{2}}^2k^2)}\,.
\end{eqnarray}
There are no minima apart from $k_\mathrm{min}=0$ within the range of applicabilty of the scattering length approximation. This is in agreement with the full numerical results shown in Fig.~\ref{fig:nd_taylor} despite the very
limited range of applicability of the scattering length approximation in nucleon-deuteron scattering.

\section{$dd$ scattering}

The general form of the scattering matrix for $s$-wave $dd$ scattering is given in Eq.~\eqref{eq:S1-1_2}. 
Due to the Bose symmetry of the two-deuteron state, 
only the $S=0$ and $S=2$ channels are present. 

\subsection{Results}

Because the $({\vec{S}_1}\cdot{\vec{S}_2})^2$ operator contributes, the $dd$ $S$-matrix is more complicated than in the previous cases and the evaluation of the entanglement entropy is computationally more expensive.
Since our investigations above demonstrate that all entanglement entropies are equally suitable for our purpose, we focus on the leading Taylor expansion of the von Neumann entropy in the $dd$ case.
The analytical expression for $\epsilon_1$ is given in Table~\ref{tab:dd_taylor}. We give the result for $s$-wave $dd$ scattering based on Eq.~\eqref{eq:S1-1_2} and, for completeness, the general result including $\delta_1$ for 
$s$-wave scattering of distinguishable spin-$1$ particles
based on Eq.~\eqref{eq:S1-1}. In the $dd$ case, the entanglement power starts with a non-zero value at $k=0$. 
This can be understood from the structure of the spin-projected initial state $\hat{P}\ket{\psi_\textrm{in}}$. The initial product state $\ket{\psi_\textrm{in}}$ contains all spins. Applying $\hat{P}$ from Eq.~\eqref{eq:defP} to project on the $S=0,2$ components, however, creates an entangled state. As a consequence, the initial state has a non-vanishing entanglement entropy, which leads to an offset in the entanglement power.
For the discussion of universal features of the entanglement created in the scattering process, only the relative maxima and minima of $\epsilon_1$ are relevant.
%%%%%%%%%%%%%%%%%%%%%%%%%%%%%%%%%%%%%%%%%%%%%%%%%%%%%%%%%%%%%%%%%%%%%%%%%%%%%%%%%%%%%%%%%%%%%%%%%%%%%%%%
\begin{table}[h]
    \caption{$s$-wave scattering contributions to the entanglement power $\epsilon_1$ for $dd$ scattering expressed through the phase shifts $\delta_{0}$ and $\delta_{2}$ for $S=0$ and $S=2$ (second line). The third line gives the general result including $\delta_1$ for the $S=1$ channel
    which applies to distinguishable particles.
    }
    \centering
    \begin{tabular}{c|l}
    \toprule
         system & $\epsilon_1$
         \\
         \midrule
         $dd$  & $\frac{1}{576}(153-70\cos[2(\delta_0-\delta_2)]-20\cos[4(\delta_0-\delta_2)])$ \\[.5ex]
         \midrule
         distinguishable & $\frac{1}{648}\Big(156-6\cos[4(\delta_0-\delta_1)]-65\cos[2(\delta_0-\delta_2)]-10\cos[4(\delta_0-\delta_2)]$\\
         particles
          & $-60\cos[4(\delta_1-\delta_2)]-15\cos(2[\delta_0-2\delta_1+\delta_2])\Big)$\\
          \bottomrule
    \end{tabular}
    \label{tab:dd_taylor}
\end{table}
%%%%%%%%%%%%%%%%%%%%%%%%%%%%%%%%%%%%%%%%%%%%%%%%%%%%%%%%%%%%%%%%%%%%%%%%%%%%%%%%%%%%
In both cases, $dd$ scattering and distinguishable particles, the entanglement power depends only 
on the difference of phase shifts for different total spins $S$. For $dd$ scattering, the pure Coulomb contribution thus cancels out in the entanglement power.\footnote{Note that this cancellation would not occur in $dd$ scattering if Eq.~\eqref{eq:S1-1} with $\delta_1=0$ was used for the $S$-Matrix instead of Eq.~\eqref{eq:S1-1_2}. In this case, the corresponding entanglement power would depend on the screening radius for the Coulomb potential and the screening could not be removed at the end of the calculation.}
As a consequence, the entanglement power for $s$-wave $dd$ scattering is determined by the Coulomb-modified strong phase shift alone, similar to the $pd$ case.

\begin{figure}[h]
\centering
\includegraphics[width=.65\textwidth]{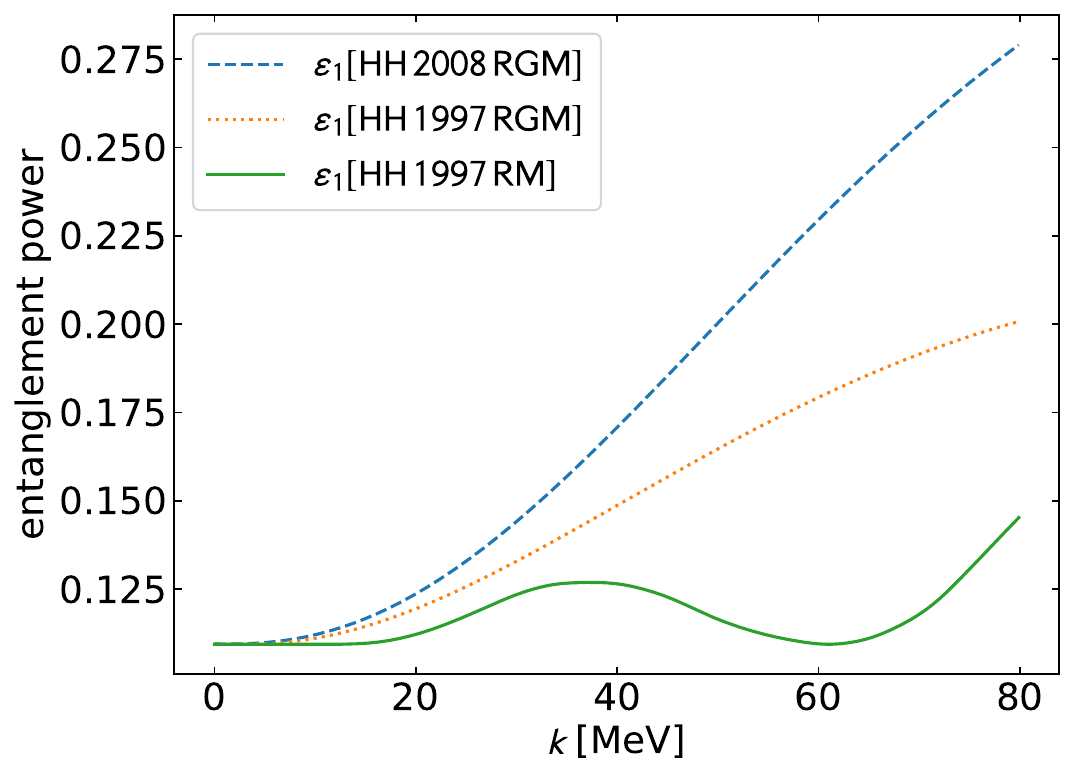}
 \caption{Entanglement power $\epsilon_1$ for $dd$ scattering phase shifts obtained by Hofmann and Hale. The solid line is based on the $R$-matrix analysis of Ref.~\cite{Hofmann:1996jv}. The dashed and dotted lines are calculated using the resonating group model (RGM) calculations for the AV18 + UIX and Bonn potentials from Refs.~\cite{Hofmann:1996jv} and \cite{Hofmann:2005iy}, respectively.   
}
\label{fig:deuterondeuteron}
\end{figure}

Next, we evaluate the entanglement power using the Coulomb-modified $dd$ phase shifts 
obtained by Hofmann and Hale~\cite{Hofmann:1996jv,Hofmann:2005iy}. In Ref.~\cite{Hofmann:1996jv}, they 
presented a calculation of $dd$ scattering in the resonating group model (RGM) for the Bonn potential and compared to a charge-independent, Coulomb corrected $R$-matrix analysis of reaction data in the four-nucleon system. This work was updated in Ref.~\cite{Hofmann:2005iy} with an RGM calculation using the AV18 
two-nucleon potential and an Urbana-IX three-nucleon force and a
new $R$-matrix analysis. As in the previous section, we evaluate the entanglement power $\epsilon_1$ using the Coulomb-modified nuclear
phase shifts.
In Fig.~\ref{fig:deuterondeuteron}, we show the corresponding results for $\epsilon_1$. 
%%%%%%%%%%%%%%%%%%%%%%%%%%%%%%%%%%%%%%%%%%%%%%%%%%%%%%%%%%%%%%%%%%%%%%%%%%%%%%%%%%%%%%%

%%%%%%%%%%%%%%%%%%%%%%%%%%%%%%%%%%%%%%%%%%%%%%%%%%%%%%%%%%%%%%%%%%%%%%%%%%%%%%%%%%%%

While the entanglement powers shown in Fig.~\ref{fig:deuterondeuteron}
agree at low momenta up to about 10~MeV, there are significant differences at higher momenta.
The RGM calculations for both potentials show a monotonic increase of $\epsilon_1$, but differ in their absolute size at larger  momenta. No universal features are evident. The $R$-matrix analysis from Ref.~\cite{Hofmann:1996jv}, however,
shows a minimum of the entanglement  power around $k=60$~MeV. The significance of this 
feature and the reason for its absence in the RGM calculations deserve further study.

Moreover, it would be interesting to investigate the entanglement power in the threshold region more closely. This will shed some light on the signature of the $^4$He excited state slightly above the scattering threshold in the
entanglement power \cite{Konig:2016utl,Gattobigio:2023fmo}. The exact location of this resonance has received some recent interest in the context of investigations of the monopole transition form factor of the $^4$He nucleus \cite{Kegel:2021jrh,Michel:2023ley,Meissner:2023cvo}.

\section{Summary}
In this paper, we have investigated the spin entanglement in few-nucleon scattering processes involving nucleons and deuterons
using the entanglement power introduced by Beane et al.~\cite{Beane:2018oxh}.
We have considered different entanglement entropies as a basis 
for the calculation of the entanglement power for the cases of spin-$\nicefrac{1}{2}$-spin-$\nicefrac{1}{2}$, spin-$\nicefrac{1}{2}$-spin-$1$ and spin-$1$-spin-$1$ scattering.
The entanglement powers were evaluated for neutron-proton, neutron-deuteron, proton-deuteron, and deuteron-deuteron scattering, taking into account the Coulomb contribution for the latter two processes. For all systems considered, the different entropies give the same information about the entanglement and are therefore equally well suited for quantifying these properties. In practice, it is 
preferable to use the original entanglement power defined in 
Ref.~\cite{Beane:2018oxh} based on the first order
Taylor expansion of the von Neumann entropy. While the linear 
approximation may in principle miss information if the reduced
density matrix is not close to the unit operator, this was not found to be the case for the considered processes.
In all considered cases, the entanglement power for $s$-wave scattering only depends on the difference of phase shifts for different spin channels.
For charged particles, the pure Coulomb contribution thus cancels out and the entanglement power is determined by the Coulomb-modified strong phase shift alone.

Finally, no universal low-energy features in the entanglement powers for neutron-deuteron and proton-deuteron scattering could be identified.  The deu\-te\-ron-deu\-te\-ron case deserves further study both in the threshold region where the $^4$He excited state resides and at intermediate momenta. In the future, it would be interesting to go beyond pure spin entanglement and 
investigate the possible manifestation of large-scattering-length universality in the spatial entanglement of light nuclear systems.

\begin{acknowledgements}
We thank Jared Vanasse for sharing the results of his pionless effective field theory calculations of neutron-deuteron scattering and Matthias Göbel for constructive comments.
This work was supported by the Deutsche Forschungsgemeinschaft (DFG, German Research Foundation) - Projektnummer 279384907 - SFB 1245 and by the German Federal Ministry of Education
and Research (BMBF) (Grant No. 05P21RDFNB).
\end{acknowledgements}

\bibliographystyle{elsarticle-num}
\bibliography{literature}   % name your BibTeX data base

\end{document}